\documentclass[runningheads,a4paper,11pt,dvipsname]{llncs}
\usepackage{eso-pic}
\AddToShipoutPicture*{
\put(133,160){
\scriptsize
Proceedings of WLPE 2013, {\tt arXiv:1308.2055}, August 2013.
}}

\usepackage{ucs}
\usepackage[utf8]{inputenc}

\usepackage{fourier}
\usepackage[T1]{fontenc}

\DeclareMathAlphabet{\mathtt}{T1}{txtt}{m}{}

\usepackage[english]{babel}

\usepackage{paralist}

\usepackage{listings}
\usepackage{algorithmicx}
\usepackage{algpseudocode}

\usepackage{graphicx}

\newcommand\angled[1]{\ensuremath{\left\langle#1\right\rangle}}

\newcommand\nonterm[1]{{\small\it#1\,}}
\newcommand\repeatz[1]{[#1]$^\ast$}

\newcommand\pto{\mathrel{\ooalign{\hfil$\mapstochar$\hfil\cr$\to$\cr}}}

\titlerunning{Implementing CHR as a Domain-Specific
  Language Embedded in Java}

\title{Implementing Constraint Handling Rules as a Domain-Specific
  Language Embedded in Java%
  \thanks{The research leading to these results has received funding
    from the European Unions Seventh Framework Programme
    (FP7/2007-2013) under grant agreement N$^{\mathrm{o}}$ 258862 4CaaSt.
  }
}

\author{Dragan Ivanovi\'c}
\institute{%
  IMDEA Software Institute\\%
  Madrid, Spain\\%
  \texttt{dragan.ivanovic@imdea.org}%
}

\begin{document}

\maketitle

\begin{abstract}
  Programming languages and techniques based on logic and constraints,
  such as the Constraint Handling Rules (CHR), can support many common
  programming tasks that can be expressed in the form of a search for
  feasible or optimal solutions.  Developing new constraint solvers
  using CHR is especially interesting in configuration management for
  large scale, distributed and dynamic cloud applications, where
  dynamic configuration and component selection is an integral part of
  the programming environment.  Writing CHR-style constraint solvers
  in a domain-specific language which is a subset of Java -- instead
  of using a separate language layer -- solves many integration,
  development cycle disruption, testing and debugging problems that
  discourage or make difficult the adoption of the CHR-based approach
  in the mainstream programming environments.  Besides, the prototype
  implementation exposes a well-defined API that supports
  transactional store behavior, safe termination, and debugging via
  event notifications.  \smallskip
\end{abstract}

\section{Introduction}
\label{sec:introduction}

Programming languages and techniques based on logic and constraints
\cite{intro_constraints_stuckey} provide programmers with powerful
high-level, declarative abstractions 
that are well suited for a wide spectrum of applications
where the computational problem can be represented as a search for
some, all, or an optimal solution (i.e., a model) that satisfies a set
of logical formulas and constraints on variables \cite{Dechter03Constraints,Apt03}.
%
%
Over time, several ecosystems of such languages, tools and programming
practices have evolved, each with a slightly different focus and
features, better suited or more specialized for one application area
or another.  Prolog
\cite{SterlingShapiro94,bratko-short,hermenegildo10:ciao-design-tplp-tr}
is probably the basis for the best known family of the Constraint
Logic Programming (CLP) language implementations, and has influenced
many others, such as Mercury \cite{mercury-manual}, Oz
\cite{mozart-oz-tutorial}, and Erlang \cite{erlang}.

In this paper, we are concerned with replicating and reimplementing
the essential features of the Constraint Handling Rules (CHR)
\cite{fruehwirth09:CHR_book}, a language that has been developed for
writing constraint solvers -- i.e., the CLP tools themselves.  CHR is
actually a rule language layer on top of a host language.  While in
principle the choice of the host language is not restrictive, the
reference implementation of CHR (and the most of the current CHR code)
works on top of Prolog \cite{schrijvers2005constraint}.  However,
there is nothing intrinsically dependent on Prolog in the semantics of
CHR rules.  Several CHR systems have been implemented on top of Java,
C and Haskell \cite{kuleuven-chr-impl}.

While there are arguably many situations where the developers using
the mainstream programming environments and tools, such as those for
Java, would benefit from using CHR techniques for writing custom
constraint solvers, developing the CHR code together with the ``main''
application / library code is still a difficult and cumbersome
process.  Even if the CHR host language is the same as the main
application language (e.g., Java), this still calls for additional
intermediate compilation tools and steps, frequently disrupts the
normal development workflow, and offers little if any rule debugging.
These practical problems, on both unit and integration level, often
discourage the use of CHR (and CLP) based techniques in the mainstream
programming environments -- ironically exactly for the problems for
which these approaches are best suited.

We argue that an effective way to address most of these problems is to
express the declarative CHR-based solver logic directly in the host
language -- in this case Java -- without introducing an additional
language layer and the intermediate compilation tools and steps.  In
the proposed approach, the CHR-based code is written in a
domain-specific language (DSL) which is a subset of Java, and the key
constraint handling components are exposed as Java objects with
well-defined interfaces that support transactional behavior, event
notifications, tracing and debugging.
The paper is based on an implementation of the proposed system.

In the remainder of the paper, we give a motivating example in
Section~\ref{sec:motivating-example}, present the DSL for the
constraint handlers in Sections~\ref{sec:chr-as-java}
and~\ref{sec:lazy-compilation-dsl}, and briefly explain their
semantics in Section~\ref{sec:semantics}.
%
%
Section~\ref{sec:impl-notes-advanc} presents some implementation
notes with the advanced transactional, debugging, and safe termination
features.  Finally, Section~\ref{sec:conclusions} offers some
conclusions.

\section{Motivating Example}
\label{sec:motivating-example}

Configuration management is one of the traditional CLP fields of
application, starting from the early systems where it was used for
querying a (static) database of available components, planning
installation steps and building the dependency chains
\cite{Dart:1991:CCM:111062.111063,amos-cbd}, to automatic
configuration of autonomous network devices \cite{5188808}, to
automated synthesis of complex components that meet some functional
requirements \cite{DBLP:conf/ijcai/PistoreMBT05}.

%
%
A significant part of the effort to build workable cloud application
platforms is related to configuration management, and relies on rich
cloud software component models \cite{DBLP:conf/sefm/CosmoZZ12}.
Additional complexities in the cloud configuration management include:

\begin{compactitem}
\item Components that are controlled and hosted by third parties that
  publish only their interfaces and descriptions.
\item Different component granularity -- from libraries to separately
  deployed virtual machines and servers.
\item Multiple configurations for coarse-grained deployable components
  that implement the same or a similar functionality.
\item Quality of Service (QoS) attributes and requirements, related to
  performance, cost, availability, and other quality concerns.
\end{compactitem}

In many cases, these aspects can be naturally addressed using
constraint models that involve not only the traditional Boolean,
finite and numeric domains, but also much richer and extensible ones.
For instance, QoS values and their ranges can be quantified using a
variety of floating point, fixed or arbitrary precision numbers with
units of measurement attached.  QoS value distributions as
mathematical objects can be represented using data sets or analytic
functions.  Regular expressions can be used to restrict service
identifiers and attributes.  Textual version information can be
converted into objects that keep hierarchic version numbering,
time-stamps and release tags.  Subsumption and compatibility
constraints can be also placed on service interfaces based on their
operations, argument and return types.

This clearly calls for constraint-solving capabilities as a part of
the runtime cloud programming environment
\cite{DBLP:conf/esocc/PredaGGMM12}.  
For most of the rich constraint domains mentioned above, there are
well tested libraries and optimized algorithms already in place, and
the object themselves are accessed through their interfaces, without
looking at the data structure implementation.
Therefore, from the interoperability point of view, the constraint
solving components should ideally behave as the standard host language
-- e.g. Java -- components, which are packaged and deployed in the
standard way, as \emph{.jar} libraries, OSGi
  bundles, or Web/application server packages.

Obviously, that is difficult to achieve if the constraint solver
implementation language is different from the host language.  But even
where that is not the case, the current limitations and maturity
levels of the systems that compile CHR into Java (e.g. JCHR
\cite{vanWeert+2005}), call for a simpler solutions which are more
closely integrated with Java.

\section{Constraint Rules as a Java DSL}
\label{sec:chr-as-java}

CHR units that implement constraint solving functionality over some
domain are called \emph{handlers}, and contain constraint declarations
and rule definitions, typically written in a CHR-specific syntax,
which admits a subset of the host language expressions and data/object
notation.
One problem with that approach is the need to translate the rules from
the CHR-specific syntax into the host language, before integrating it
with the rest of the application and libraries.  When testing and
debugging, it can be difficult to trace the solver behavior back to
the CHR source code.  Another problem is that the CHR syntax has to be
updated from time to time to keep up with the innovations in the host
language, such as the introduction of generics and enumerations in
Java 5, enhanced type inference in Java 7, or the forthcoming
introduction of lam\-bda-expressions in Java 8.  While these new
features normally do not deprecate the old ones, keeping up-to-date
is certainly desirable.%
\footnote{This is not just a matter of experimental features.  In a
  language like Java, where each language innovation comes after a
  prolonged process of drafting and discussion, inclusion of a new
  language feature usually means that most coders will start using it
  very soon.}

To simplify and streamline the integration with the host language, we
propose to express handlers, constraints, and rules in a
domain-specific language which is a subset of Java.  This is not
unlike the inversion-of-control design pattern: instead of the CHR
level controlling Java classes, we let Java code construct and
configure CHR handlers using specific APIs.  Instead of a static
CHR-to-Java compilation, we use a transparent runtime compilation of
the handler logic into the back-end Java objects that fire rules and
update the store.  And instead of imposing restrictions on Java
constructs that are recognized by CHR, we define CHR-specific APIs
callable from arbitrary Java code.

\begin{figure}[tb]
\centering
\begin{minipage}[t]{0.6\textwidth}
\begin{lstlisting}
import cr.core.Symbol;
import cr.core.Handler;
// Other imports

public class MyHandler<...> extends Handler {

  // Symbol declarations

  public MyHandler(...) {
    // Constructor code
    initialize();
  }

  public void setup() {
    // Constraint declarations
    // Constraint rules
  }

  // Guard methods
  // Other methods
}
\end{lstlisting}
\end{minipage}
\caption{The general form of a constraint handler.}
\label{fig:template}
\end{figure}

Figure \ref{fig:template} shows the general shape of a constraint
handler in our approach.  Each handler class, which may have type
parameters, extends the abstract class \texttt{cr.core.Handler}, which
is the part of the CHR-in-Java library.  The other imported class,
\texttt{cr.core.Symbol} is used to name constraints and data
elements.  The four main DSL-specific parts of the handler are: symbol
declarations, constraint declarations, rule definitions, and guard methods.

\begin{figure}[tb]
  \centering
  \begin{minipage}[t]{0.75\textwidth}
    \tt\footnotesize
    \begin{tabbing}
      \nonterm{SymbolDecl} ::= \=\textbf{public} Symbol \nonterm{Name}[,
      \nonterm{Name}]$^\ast$ ;\\
      \>\nonterm{Name} ::=\'\angled{\mbox{\it a valid Java field name}}
    \end{tabbing}
  \end{minipage}
  \caption{Syntax for symbol declarations.}
  \label{fig:symdecl}
\end{figure}

Symbol declarations follow the simple scheme from
Figure~\ref{fig:symdecl}.  Two predefined symbols in
\texttt{cr.core.Handler} are \texttt{fail} (representing the
unsatisfiable constraint) and \texttt{\char95} (the underscore, used
to represent an arbitrary object).  Note that the symbol fields are
public, but not initialized -- the \texttt{initialize()} method of
\texttt{cr.core.Handler} -- which needs to be called at the end of a
custom handler constructor -- uses Java reflection to initialize each
public field of type \texttt{cr.core.Symbol} to a fresh symbol with
the same name.  No particular naming strategy is enforced, but it is
customary to use names starting with a lowercase letter for
constraints, and those starting with an uppercase letter for data
objects.

One advantage of declaring symbols using public fields is that one can
use the usual Java refactoring tools in modern IDEs, such as Eclipse,
NetBeans, or IntelliJ/IDEA, to perform project-wide consistent
renaming of constraints.

\begin{figure}[tb]
  \centering
  \begin{minipage}[t]{0.75\textwidth}
    \tt\footnotesize
    \begin{tabbing}
      \qquad~\nonterm{Decl} ::= 
      \=
      constraint(\nonterm{Name}[, \nonterm{KeyClass}]$^\ast$)
      \\
      \>
      [.with(\nonterm{DataClass}[, \nonterm{DataClass}]$^\ast$)] ;
      \\[3pt]
      \>
      \nonterm{KeyClass} ::=%
      \'%
      \angled{\mbox{\it a Java expression of type~~\rm\tt
          Class<Comparable>}}
      \\
      \>
      \nonterm{DataClass} ::=%
      \'
      \angled{\mbox{\it a Java expression of type~~\rm\tt Class<?>}}
    \end{tabbing}
  \end{minipage}
  \caption{Syntax for constraint declarations.}
  \label{fig:const-decl}
\end{figure}

The handler class needs to implement method \texttt{setup()} which is
called from \texttt{initialize()}, whose task is to declare the
constraints and define the rules.  Constraint need to be declared
before being used in a rule, and figure \ref{fig:const-decl} shows the
corresponding DSL syntax.  Each constraint is uniquely identified with
its \textit{Name} (a declared symbol), and may have zero or more key
and data fields, whose classes are given in the declaration.  The key
fields hold \texttt{Comparable} objects, and they uniquely identify
a constraint literal instance, while the data fields (introduced after
``\texttt{.with}'') carry additional information (arbitrary objects)
associated to the constraint literal instance, which may vary over
time.  Nulls are allowed in both the key and data fields.

For instance, the following statements in \texttt{setup()}:
\begin{lstlisting}
constraint(leq, String.class, String.class); // less-than-or-equal
constraint(lt, String.class, String.class);  // less-than
constraint(eq, String.class, String.class);  // equal
constraint(neq, String.class, String.class);  // not equal
\end{lstlisting}
declare constraints named \texttt{leq}, \texttt{lt}, \texttt{eq}, and
\texttt{neq} (all declared symbols) between two string keys
(constrained variable names).  Also:
\begin{lstlisting}
constraint(dom, String.class).with(Integer.class, Integer.class);
\end{lstlisting}
declares a constraint \texttt{dom} which associates a range of integer
values (between the limits in the data fields) to a variable whose
name is given as the key.

\begin{figure}[tb]
  \centering
  \begin{minipage}[t]{0.9\textwidth}
    \tt\footnotesize
    \begin{tabbing}
      \qquad\qquad\qquad\qquad\llap{\nonterm{Rule} ::=}
      \=
      \nonterm{Head} [\nonterm{Guard}] [\nonterm{Body}] ;
      \\[8pt]
      \>
      \nonterm{Head} ::=%
      \'
      when(\nonterm{Name}[, \nonterm{Pattern}]) 
      [.with(\nonterm{Pattern})]
      \nonterm{Modifiers}
      \\
      \>
      \repeatz{.and(\nonterm{Name}[, \nonterm{Pattern}]) 
        [.with(\nonterm{Pattern})]
        \nonterm{Modifiers}}
      \\[3pt]
      \>
      \nonterm{Guard} ::=%
      \'
      .where(\nonterm{GuardName}[, \nonterm{Pattern}])\\
      \>\repeatz{.and(\nonterm{GuardName}[, \nonterm{Pattern}])}
      \\[3pt]
      \>
      \nonterm{Body} ::=%
      \'
      .then(\nonterm{Name}[, \nonterm{Pattern}])
      \\
      \>
      \repeatz{.and(\nonterm{Name}[, \nonterm{Pattern}])}
      \\[6pt]
      \>
      \nonterm{Pattern} ::=%
      \'
      \angled{\mbox{\it a Java expression}}
      [, \nonterm{Pattern}]
      \\
      \>
      \nonterm{Modifiers} ::=%
      \'
      [.passive()] [.keep()]
    \end{tabbing}
  \end{minipage}
  \caption{Syntax for rule definitions.}
  \label{fig:rule-def}
\end{figure}

The syntax for rules is more complex, and is given in
Figure~\ref{fig:rule-def}.  In this section we present different
aspects of the rule definitions with an informal explanation of their
intended meaning.  More detailed discussion of the rule semantics is
given in the Section~\ref{sec:semantics}.

The simplest rule may have only a head, as
in the following two examples:
\begin{lstlisting}
when(leq, X, X);
when(eq, X, X);
\end{lstlisting}
which (with \texttt{X} a declared symbol) simply consume or throw away
the trivial (in)equalities.  The fields are compared on the basis of
the \texttt{equals()} method.

Most often, rules have a body.  An example of a simplification
rule is:
\begin{lstlisting}
when(lt, X, Y).then(leq, X, Y).and(neq, X, Y);
\end{lstlisting}
which converts strict inequality $x<y$ into the equivalent conjunction
of $x\leq y$ and $x\neq y$.  Another simplification example is:
\begin{lstlisting}
when(neq, X, X).then(fail);
\end{lstlisting}
which detects inconsistencies.  An example of \texttt{.passive()}
modifier is:
\begin{lstlisting}
when(leq, X, Y)
.and(leq, Y, X).passive()
.then(eq, X, Y);
\end{lstlisting}
which simplifies $x \leq y \land y \leq x$ into $x=y$, but since the
case is completely symmetric, wants to avoid firing twice, on $y\leq
x$.  Another use of \texttt{.passive()} is to prevent proliferation of
non-informative facts.  For instance:
\begin{lstlisting}
when(eq, X, Y)
.and(eq, X, Y).passive().keep();
\end{lstlisting}
consumes $x=y$ if that fact is already known.  Note also the
modifier \texttt{.keep()} which prevents the known fact from being
consumed too.

In fact, modifier \texttt{.keep()} is the mechanism for implementing
propagation and simpagation rules.  For instance, the following rule
ensures the symmetry of \texttt{eq}:
\begin{lstlisting}
when(eq, X, Y).keep()
.then(eq, Y, X);
\end{lstlisting}
and the next one propagates the domains of the equal variables:
\begin{lstlisting}
when(eq, X, Y).keep()
  .and(dom, X).with(A, B).keep()
  .and(dom, Y).with(C, D).keep()		
  .where("!equals", X, Y)  // avoid the trivial case
  .then(dom, X).with(C, D)
  .and(dom, Y).with(A, B);
\end{lstlisting}

In the last example, we have seen an example of a guard, introduced
with ``\texttt{.where}'', whose first argument is a string that points
to the corresponding guard method, with the leading bang
(``\texttt{!}'') signifying the negation.  The corresponding guard
method:
\begin{lstlisting}
public boolean equalsGuard(Object x, Object y) {
  return (x == null ? y == null : y != null && x.equals(y));
}
\end{lstlisting}
is built into \texttt{cr.core.Handler}.  A non-negated guard succeeds
when all of the arguments have the correct type, and the returned
value is \texttt{true} (or the guard method return type is
\texttt{void}).  A negated guard succeeds exactly when the non-negated
guard would fail.

The guard mechanism is very flexible and powerful, handles the
automatic conversion between Java primitive values and objects, accepts
variable argument lists, and allows the guard methods to compute new
information that can be used in the body of the rule.

For instance, the following rule detects inconsistencies:
\begin{lstlisting}
when(dom, X).with(A, B)
.and("!lessOrEqual", A, B)
.then(fail);
\end{lstlisting}
using the rule method:
\begin{lstlisting}
public boolean lessOrEqualGuard(int a, int b) {
  return a <= b;
}
\end{lstlisting}
(Note that \texttt{"!lessOrEqual"} guard succeeds  if either of
the two arguments are \texttt{null}.)

This rule ignores non-informative bounds:
\begin{lstlisting}
when(dom, X).with(A, B)                   // newly told
.and(dom, X).with(C, D).passive().keep()  // already known, kept
.where("includes", A, B, C, D);           // [C,D] already included in [A,B]
\end{lstlisting}
with the guard method:
\begin{lstlisting}
public boolean includesGuard(int a, int b, int c, int d) {
  return (a <= c) && (d <= b);
}
\end{lstlisting}

The following rule treats the informative bounds:
\begin{lstlisting}
when(dom, X).with(A, B)  // newly told
.and(dom, X).with(C, D).passive() // already known
.where("!includes", A, B, C, D) // [A,B] does not include [C,D]
.and("isect", A, B, C, D, E, F) // [E,F] is the intersection
.then(dom, X).with(E, F); // update the domain to [E,F]
\end{lstlisting}
with the new guard method that computes the intersection:
\begin{lstlisting}
public void isectGuard(int a, int b, int c, int d, 
               @NotNull Symbol e, @NotNull Symbol f) {
  e.set(Math.max(a, c));
  f.set(Math.min(b, d));
}
\end{lstlisting}
This guard always succeeds (if no argument is \texttt{null}), and
stores the results in \texttt{e} and \texttt{f}, used in the rule body
as the updated value range.  Guard method parameters of type
\texttt{cr.core.Symbol} are passed not by value, but by reference.


\section{Runtime rule compilation and DSL expressiveness}
\label{sec:lazy-compilation-dsl}

At this point, before proceeding to the semantics, it is useful to
comment on some aspects of the proposed approach and its
implementation, and to highlight and motivate the choices these are
based on.

First and foremost, the elimination of the static rule compilation
phase, as mentioned at the beginning of the previous section, comes at
the cost of a runtime compilation of rules into Java objects.  In the
current implementation, this is done every time a new instance of the
handler is instantiated (i.e., during and after the execution of the
\texttt{setup()} method), but in a slightly improved implementation,
most of this overhead can be dealt with on once-per-class basis,
provided that \texttt{setup()} does not depend on the handler
constructor parameters.

The first runtime rule compilation phase is building the internal rule
object representations, which is done using the \texttt{constraint()},
\texttt{when()}, \texttt{where()}, \texttt{then()}, and other API
methods.  The most complex part here is the treatment of guards, which
relies on Java reflection to ensure that the corresponding methods
exist, and to create adapters that take care of the correct argument
count, types, conversion, variable-argument list passing, returning
result interpretation, etc.
The use of strings for guard names, while not as elegant as the other
parts of the DSL, allows the use of the negation prefix
(``\texttt{!}'') and avoids the need to declare guard names as
symbols, and thus clutter the code.%
\footnote{Note that Java method names and field names populate
  different namespaces.}
The second runtime compilation phase is weaving the compiled rules
into a per-instance index structure that is used for firing rule
heads.

A handler instance can be created only if the runtime rule compilation
succeeds.  Otherwise, a \texttt{cr.core.HandlerException} is thrown
with a fault description.  Not having all the errors in the handler
detected statically is arguably the greatest drawback of our scheme,
although it is less critical in the context of the agile development
methodologies.  Any runtime rule compilation errors would be weeded
out early on during the handler's unit testing phase, before
integrating it with the rest of the application modules.

It should also be noted that JVM-based languages such as Scala
\cite{scala-lang} provide much better facilities for development of
DSLs than ``pure'' Java.  In particular, Scala's flexible system for
defining operators, together with a functional representation of
methods as first-class objects (on the same level as the variable and
value fields) may eliminate the need for strings as guard names and
run-time argument number and type checking.  This makes implementing a
Scala interface for the constraint rules library an interesting
next step.  Scala can be also used as the implementation language, but
since it introduces its own object (reference / value) hierarchy on
top of Java, this would be be more suitable when the client code is
also written in Scala.

\section{Semantics}
\label{sec:semantics}

The semantics of the constraint rules introduced in
Section~\ref{sec:chr-as-java} as a Java DSL follows the general lines
of CHR, but differs from its standard semantics with respect to the
organization of the store, the firing of rules, and the absence of
special built-in solvers.
%
%
Each instance of the handler class (i.e., the one that extends
\texttt{cr.core.Handler}) encapsulates five key elements: the symbols,
the store, the goal (or the queue), and the rules, which are explained
below.

As mentioned in the previous section, symbols are just objects with an
immutable name, and are used to name the constraints and data field
values in rules.  (Using the same symbol for both purposes is not
forbidden, but the resulting code may look confusing.)  When testing
guards and firing rules, symbols denoting data fields also store field
values as objects.  Since the rules operate on the committed-choice
basis, this is done using destructive updates, by calling their
\texttt{.set()} and \texttt{.get()} methods.  When used as  data
objects outside the rules, the symbols' values should be treated as
volatile.

\sloppy
The handler state is a tuple $(G,S)$, where $G$ is the goal, and $S$
is the store.
The \emph{store} is an object that keeps the \emph{known facts} about
the declared constraints.  Unlike the standard CHR where the store is
a multi-set of constraint literals, we take the approach where each
declared constraint $c$ is a partial function of the form:
\begin{displaymath}
  c : K_1 \times K_2 \times \cdots \times K_n \pto D_1 \times D_2
  \times \cdots \times D_m
\end{displaymath}
where $n,m\geq 0$. 
Each $K_i$ corresponds to a Java class implementing
\texttt{java.lang.Comparable} interface, and each $D_j$ to an
arbitrary Java class.
(Each $K_i$ and $D_j$ is also implicitly extended to include the null
reference.)  If $n$ or $m$ is zero, the corresponding product
degenerates to a singleton set containing only the unit tuple $()$.  Each
constraint literal (or \emph{fact}) is a statement of the form:
\begin{displaymath}
  c : (k_1,k_2,\ldots,k_n) \mapsto (d_1,d_2,\ldots,d_m)
\end{displaymath}
where $k_i\in K_i$ and $d_j\in D_j$, which tells that $c$ is defined
at $(k_1,k_2,\ldots,k_n)$ and has value $(d_1,d_2,\ldots,d_m)$.
Initially, the store is typically empty, which means that all
constraints are undefined for all possible keys.  
For instance, if the
\textit{dom} constraint is declared as:
\begin{displaymath}
  \mathit{dom} : \mathbb{V} \pto \mathbb{Z} \times \mathbb{Z}
\end{displaymath}
where $\mathbb{V}$ is a set of variable labels (as strings), and the
two integers are the pair of min/max bounds, then the partial function
representation ensures that we may have at most one pair of bounds in
the store for any variable label.

\fussy

The partial function representation of constraints is chosen over the
multi-sets as a more structured solution which can leverage the
efficient Java data structures such as \texttt{java.util.SortedMap},
and is more powerful than the set semantics.  Note that the set
semantics can be simulated by taking $m=0$ and keeping all constraint
data in the key fields.  Similarly, the multi-set semantics can be
simulated by taking, e.g., $n=1$ and
$K_1\equiv\mathtt{java.lang.Integer}$, putting all constraint
information into the data part, and making sure that $k_1$ is always
ignored in the \texttt{when()} part of the rules (using symbol
``\texttt{\_}''), as well as initialized to a fresh value for each new
fact inserted into the store.

In contrast to the store which contains the already known facts, the
\emph{goal} is a conjunction of \emph{newly told} facts that await
processing.  The goal is processed one fact a time, in a chronological
(or left-to-right) order, and new facts produced by firing rules are
appended to it.  For these reasons, the goal is also known as the
\emph{queue}.

Note that our proposal does not make the distinction between the
\emph{built-in} and \emph{user-defined} (relational) constraints.  All
constraints used in the solver have to be declared, and their rules
explicitly specified.%
\footnote{This does not mean writing huge monolitic solvers.  The
  developers can use subclassing, delegation and other usual Java
  techniques for software modularization and reuse.}
Also, any object inspection and matching has to be done explicitly by
invoking the accessor methods in guards.

\begin{figure}[tb]
  \begin{minipage}[t]{0.40\linewidth}
    \begin{algorithmic}[1]
      \Function{MainLoop}{} \State $\mathit{forcedExit \gets
        \mathtt{false}}$ \While{$\neg\mathit{forcedExit} \land |G|>0$}
      \State $\phi \gets \mathit{first}(G)$; $G \gets
      \mathit{rest}(G)$ \If{$\phi\equiv \mathtt{fail} : () \mapsto
        ()$} \State $\mathit{signal\ failure}$ \Else \State
      \Call{FireAllRules}{$\phi$} \If{$|G| > \mathit{limit}$} \State $\mathit{forcedExit \gets \mathtt{true}}$
      \EndIf
      \EndIf
      \EndWhile
      \State\Return{$|G|>0$}
      \EndFunction
      \State
      \Function{Tell}{$\phi$}
        \State {\it append $\phi$ to $G$}
        \State \Return{\Call{MainLoop}{}}
      \EndFunction
    \end{algorithmic}
  \end{minipage}%
  \begin{minipage}[t]{0.58\linewidth}
    \begin{algorithmic}[1]
      \Procedure{FireAllRules}{$\phi$} \State $U\gets\emptyset$
      \For{$\mathit{each\ active\ head\ element}\ H\ \mathit{matching}\ \phi$} 
        \State $\bar H' \gets \mathit{all\ head\ elements\ in\ the\ rule\ except}\ H$ \State $\mathit{fired} \gets \mathtt{false}$
      \ForAll{$\mathit{facts}\ \bar\phi'\ \mathit{from}\ S\
        \mathit{matched\ by}\ \bar H'$} \If{\textit{the rule guard
          succeeds}} \State $\mathit{fired} \gets \mathtt{true}$
      \State {\it append the rule body to $G$} \State $U\gets U\cup\left\{
        \bar\phi_{i}' \:|\: \bar H_{i}' \mathit{without}\
        \mathtt{.keep()}\right\}$
      \EndIf
      \EndFor
      \If{$\mathit{fired} \land H\ \mathit{without}\
        \mathtt{.keep()}$} 
        \State $S\gets S\setminus U$ 
        \State\Return{}
      \EndIf
      \EndFor
      \State $S\gets (S\setminus U)\cup \{\phi\}$
      \EndProcedure
    \end{algorithmic}
  \end{minipage}
  \caption{The main loop and the rule firing algorithms.}
  \label{fig:main-loop}
\end{figure}

For simplicity, we present here the operational semantics of the rules
using the algorithms from Figure~\ref{fig:main-loop}, which
destructively update the state.  The \textsc{MainLoop} starts from
some initial state $(G,S)$ -- where $G$ is normally non-empty -- and
tries to reach a fixpoint state $(G',S')$ where $G'$ is empty, i.e.,
all possible rules have been fired and nothing else remains to be
done.  That is achieved by successively reading facts from the goal
(in a FIFO fashion), and firing all applicable rules (or signaling a
failure).  \textsc{MainLoop} is typically initiated with a
\textsc{Tell} operation which communicates a new fact to the handler.

\textsc{MainLoop} can return before reaching a fixpoint in two cases:
when explicitly asked to do so from a rule guard (using the
\texttt{forceExit()} handler method), or when it detects that the goal
size has exceeded some optional and pre-configured safety level set to
prevent uncontrolled memory consumption.  In both cases such an early
termination is safe, in the sense that no information is lost, and
that the computation can always be resumed.

Procedure \textsc{FireAllRules} uses an internal index structure to
iterate through all head elements of all rules that are active (i.e.,
not marked with \texttt{.passive()}).  For each such active head
element, an attempt is made to fire the rule for each combination of
facts from the store that correspond to the remaining head elements in
the same rule (for which the rule guard succeeds if present).  The
facts consumed in each firing (not marked with \texttt{.keep()}) are
marked for removal, but are not removed immediately to give chance to
all other applicable rules to fire.  If the firing head element is
marked with \texttt{.keep()}, the next firing head (in the same rule
or one of the following rules) is tried, otherwise the processed fact
is consumed, and the processing stops.  If the processed fact is not
consumed by any rule, it is added to the store.

The order of rules is significant.  The rules are fired in the same
order in which they are defined.  When an earlier rule consumes the
processed fact, it effectively cuts the remaining rules off.  However,
those rules that do fire behave as if they do so simultaneously, since
the removal of the consumed facts is performed at the end.

While the handler's \texttt{tell()} method informs it of the new
facts, which are added to the goal (i.e., the queue), the results of
the computation are held in the store, and can be inspected using the
\texttt{select()} method.


\section{Implementation Notes and Advanced Features}
\label{sec:impl-notes-advanc}

The current implementation is a set of Java classes and interfaces
packaged in a lightweight standalone \emph{.jar} file, without
external dependencies.%
\footnote{An archive with the binaries and the documentation can be
  downloaded from \texttt{http://software.imdea.org/\~{ }idragan/cr}}
It contains an example numerical interval solver that can be tested
using the visual debugger based on the advanced features below.

\subsection{Event notifications, tracing and debugging}
\label{sec:event-notif-trac}

\begin{figure}[tb]
  \centering
  \includegraphics[width=1.0\textwidth]{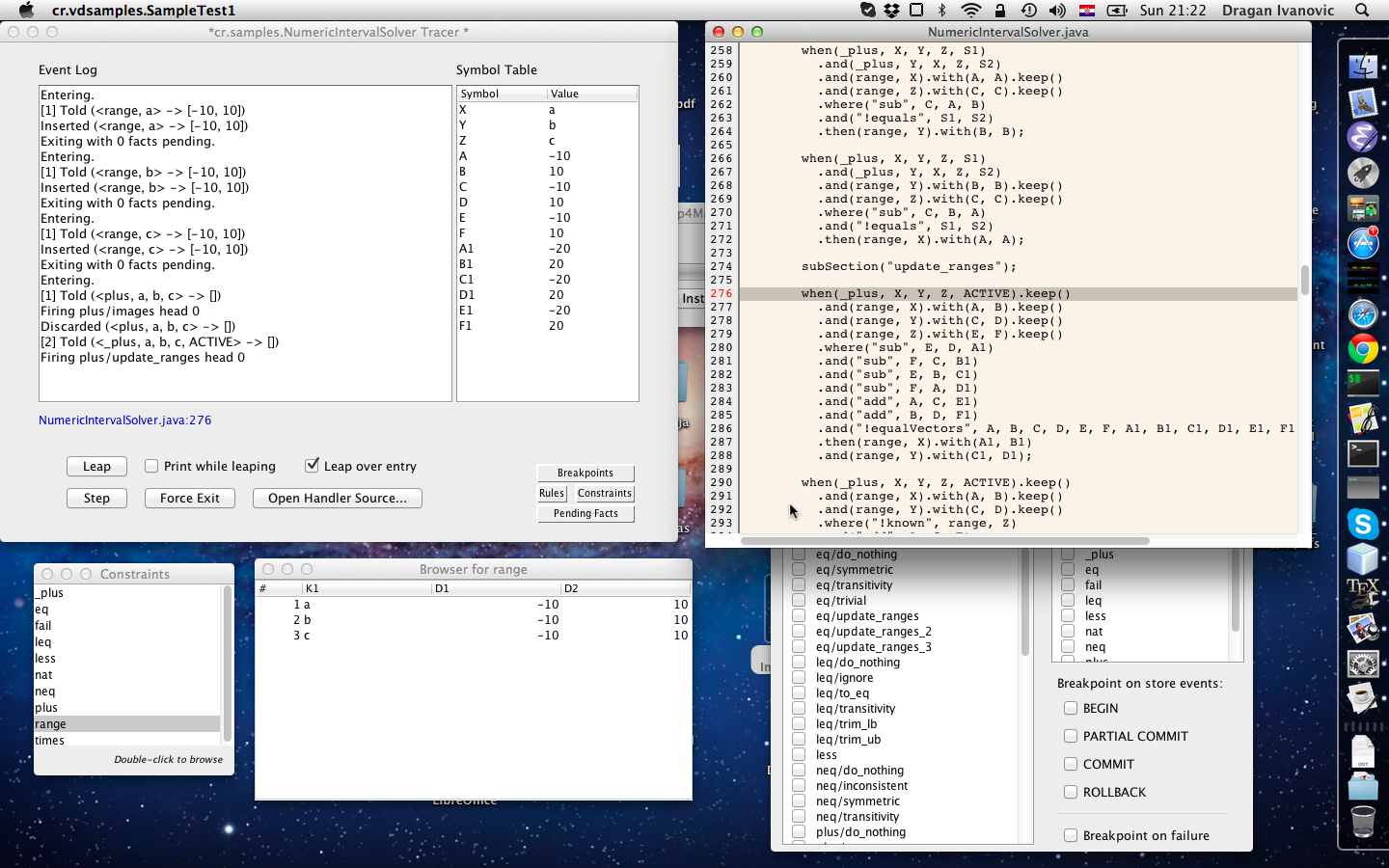}
  \caption{A screeshop of the debugging session.}
  \label{fig:debug}
\end{figure}

Insufficient support for debugging is one of the key disadvantages of
the current CHR implementations.  In our implementation, both tracing
and debugging are achieved using the publish-subscribe mechanism by
which one or more event listeners can be attached to a handler
instance and can observe different events, such as adding a goal to
the store or firing a rule.  In the latter case, the information from
Java reflection relates the firing point to the handler source.  An
example of the full GUI debugging session is given in
Figure~\ref{fig:debug}, with the debugging console, source code
tracing, breakpoints, and constraint views.

\subsection{Transactional state behavior}
\label{sec:trans-state-behav}

It is often useful to save the state of the handler before telling
more constraints, and to revert to the previous state if the problem
turns out to be over-constrained or insatisfiable.  A typical use
would be checking if a solution to the problem exists under some
additional assumptions, and if not reverting to the previous state and
trying something else.  Our implementation enables arbitrarily nested
state savepoints, analogous to those in the transactional database,
using the following operations:
\begin{itemize}
\item \texttt{begin()} -- saves the current state and begins a new,
  nested transaction.
\item \texttt{commit()} -- closes the current nested transaction and
  saves its current state to the parent transaction.
\item \texttt{partialCommit()} -- saves the current state to the
  parent transaction, while keeping the nested transaction open.
\item \texttt{rollback()} -- discards the current nested transaction
  and returns to the parent transaction and its saved state.
\end{itemize}
These operations work on both components of the state (the goal and
the store), and are orthogonal to the \texttt{tell()} and
\texttt{select()} handler operations.  The default store that is
created for each new handler instance is a map-based in-memory store.
We are working on an implementation where the in-memory store can be
replaced with a persistent store stored in the file system.

\section{Conclusions}
\label{sec:conclusions}

Implementing CHR as a domain-specific language embedded in Java has
several advantages over the classical approach where CHR handlers are
written in an additional language layer on top of the host language,
here Java.  These advantages include avoiding the additional
compilation steps that disrupt the usual development cycle, better
leverage of the host language features, support for tracing and
debugging, and the application of the existing powerful refactoring
tools in the modern Java IDEs.  On the overall, this can help improve
the acceptance of CHR and CLP programming techniques in the
component-based, Java-centric, cloud programming environment.

The future work will be directed towards more robust implementations,
integration with persistent transactional store back-ends, development
of a spectrum of ready-to-use constraint handlers, introduction of
some CHR$^\lor$ features \cite{chrd-98}, and exploring applications in
distributed event processing.

\bibliographystyle{plain}
\bibliography{all}

\end{document}